\newcommand{\ba}{\begin{array}}
\newcommand{\ea}{\end{array}}
\newcommand{\bean}{\begin{eqnarray*}}
\newcommand{\eean}{\end{eqnarray*}}
\newcommand{\rm}[1]{\mathrm{#1}}

\newcommand{\ep}{\epsilon}

\renewcommand{\d}{\partial}

\newcommand{\bra}[1]{\left\langle {#1} \right|}
\newcommand{\ket}[1]{\left|  #1 \right\rangle}

\newcommand{\braket}[2]{\left\langle #1 | #2 \right\rangle \;}

\newcommand{\bmx}[1]{\left(\begin{array}{*{#1}{c}}}
\newcommand{\emx}{\end{array}\right)}
\newcommand{\bmxw}[1]{\renewcommand{\arraystretch}{2}\left(\begin{array}{*{#1}{c}}}
\newcommand{\bmxww}[1]{\renewcommand{\arraystretch}{2.5}\left(\begin{array}{*{#1}{c}}}
\newcommand{\bdet}[1]{\renewcommand{\arraystretch}{1.2}
	\left|\begin{array}{*{#1}{c}}}
\newcommand{\edet}{\end{array}\right|\renewcommand{\arraystretch}{1}}
\newcommand{\beq}{\begin{equation}}
\newcommand{\eeq}{\end{equation}}
\newcommand{\bea}{\begin{eqnarray}}
\newcommand{\eea}{\end{eqnarray}}

\newcommand{\ditem}[1]{\item[$\diamond$]}

\newcommand{\bit}{\begin{itemize}}
\newcommand{\eit}{\end{itemize}}

\newcommand{\eab}{\begin{eqnarray}}
\newcommand{\eae}{\end{eqnarray}}

\documentclass[aps,prb,twocolumn,superscriptaddress]{revtex4-1}
\usepackage{graphics}
\usepackage{epsfig}
\usepackage{amssymb}
\usepackage{color} 
\usepackage{subfigure}

\begin{document}


\title{Static Electric Field in a 1D Systems without Boundaries}


\author{Kuang-Ting Chen}
\affiliation{Department of Physics, Massachusetts Institute of Technology, Cambridge, MA 02139}
\author{Patrick A. Lee}
\affiliation{Department of Physics, Massachusetts Institute of Technology, Cambridge, MA 02139}


\date{\today}

\begin{abstract}
In this brief report, we show that in a 1D system with unit-cell doubling, the coefficient of the $\theta$-term is not only determined the topological index, $\int i\bra{u_k}\frac{\d}{\d k}\ket{u_k}{\rm d}k$. Specifically, the relative position between the electronic orbitals and the ions also alters the coefficient. This resolves a paradox when we apply our previous result to the Su-Shreiffer-Heeger model where the two ground states related by a lattice translation have $\theta$ differed by $\pi$. We also show that the static dielectric screening is the same with or without boundaries, on the contrary to what we have commented in our previous paper.

\end{abstract}


\maketitle

\section{Introduction}
In our original paper\cite{mywork}, we argue that in a setting without boundaries, the topological insulator in one dimension (1D) and three dimensions (3D) can still be characterized by a $\theta$-term in the effective theory, which in turn gives measurable consequences. Specifically, in 1D there will be a term $(e\theta/2\pi) E$ in the effective Lagrangian, and it results in a constant electric field $\frac{\theta e}{2\pi}$ in the bulk, provided that the electric field is confined in one dimension.

This observation, however, seems puzzling when one considers the well-known Su-Schreiffer-Heeger (SSH) model\cite{ssh}. If we consider spinless electrons, the two ground states in this model will have the effective $\theta$-term with $\theta$ which differs by $\pi$. On the other hand, the two states are physically identical, and thus cannot have different ground state electric field. In this report we will resolve this issue.

Another conceptual problem is whether the electric field can be screened in a setting without boundaries. Naively speaking, one might imagine that the dielectric screening comes from the accumulated charges at the two boundaries. Without boundaries these charges are absent, and there seems to be no screening. This, however, violates the intuition that in a gapped system in 1D, boundary conditions are usually irrelevant for the bulk properties. We will show in the following that the electric field in the bulk is indeed affected by the dielectric constant. There are two ways to understand the effect: either we can claim that it is $\epsilon E$ which is quantized in integer times $e$ with a shift $-\theta e/2\pi$, or we can say that the shift in quantization of $E$ is still given by $\theta$, but the effective $\theta$ is shifted by the finite electric field it generates. In our original paper, we overlooked the effect of $\epsilon$ on the quantization condition.

In Sec. II we look into the SSH model, verify the topological index of the ground states. We then explain how we can resolve the apparent contradiction. In Sec. III we explain the dielectric screening effect in the 1D setting without boundaries.

\section{the SSH model, Topological Index, and the static electric field}
The SSH model is given by the following Hamiltonian in 1D\cite{ssh}:
\beq
H=\sum_{i,\sigma}(-t+(-1)^i\Delta)c^\dag_{i\sigma}c_{i+1\sigma}+h.c.,
\eeq
$\Delta$ takes either positive or negative values for the two ground states which spontaneously break the lattice translation symmetry. Suppose we plug in the wave function
\beq
\psi_k=a_k\sum_{i\in {\rm odd}}c^\dag_i\ket{0}\exp(ikx_i)+b_k\sum_{j\in {\rm even}}c^\dag_j\ket{0}\exp(ikx_j),
\eeq
The Hamiltonian can be put into a matrix form:
\beq
H_k\left(\begin{array}{c}a_k\\b_k\end{array} \right)
=(-2t\cos(ka)\sigma_x+2\Delta\sin(ka)\sigma_y)\left(\begin{array}{c}a_k\\b_k\end{array} \right);
\eeq
$\sigma_x$ and $\sigma_y$ are Pauli matrices and $a$ is the lattice spacing. Notice that $H_k$ is \textit{not} periodic in $\pi/a$; nevertheless $\psi_k$ is periodic (up to a phase.) When we apply a small electric field, the coupling enters via Peierls substitution, and directly results in $H_k\rightarrow H_{k+eA}$, where $A$ is the spatial part of the gauge field. At half filling where the system is insulating, following our previous discussion, we can calculate the Berry's phase accumulated when we slowly turn on the electric field until the system reaches the state related to the initial state by a large gauge transform of winding number one: (hereafter when we write "the Berry's phase" we refer to the Berry's phase of this procedure)
\beq
\theta_{\rm Berry}=\int^{\pi/2a}_{-\pi/2a}i\bra{u_k}\frac{\d}{\d k}\ket{u_k}{\rm d}k,
\eeq
with $\ket{u_k}=\left(\begin{array}{c}a_k\\b_k\end{array} \right)$, and we choose the phase such that $\psi_k$ periodic in $k$. If we take $x_n=na$, we can parametrize our solution as
\beq
\ket{u_k}=\exp\left(\frac{i{\rm sgn}(\Delta)f(k)}2(\sigma_z-1)\right)\left(\begin{array}{c}1\\1\end{array} \right),
\eeq
with 
\beq
\tan(f(k))=\left|\frac{\Delta}{t}\right|\tan(ka).
\eeq
The important thing here is to notice that $f(k)=0$ at $k=0$ and $f(k)=\pm\pi/2$ at $k=\pm\pi/2a$. We therefore get
\beq
\label{what}
\theta_{\rm Berry}={\rm sgn}(\Delta)\frac{\pi}2,
\eeq
for each spin. The coefficient $\theta$ in the $\theta$-term is just the sum of the Berry's phases.

If we consider the spinful case as in the original SSH model, the total Berry's phase differs by $2\pi$ for the two states, which implies that both would have the same ground state properties. However, since $\theta=\pi$ for both states, we thus predict that there is a electric field $E\sim \pm e/2$ in both states. If we consider the spinless case, the situation becomes even worse, as the two states are related by a lattice translation; yet they have different ground state properties.

These paradoxical observations can be resolved, if we realize that the charged ions can also have a Berry's phase. It is somewhat surprising in the sense that the ions are considered to be stationary localized charges and behave rather trivially. To see how the Berry's phase comes about, we first consider the effect of translation of the wave function on the Berry's phase:

Consider a wave function $\psi_k=u_k(x)\exp(ikx)$. (We suppress the dependence on $x$ of $u_k$ from here on when there is no ambiguity.) Let us translate the wave function by $x_0$ and denote the shifted wave function $\psi'$: 
\bea
\psi_k'&=&u_k\exp(ik(x-x_0))\nonumber\\
&=&(u_k(x-x_0)\exp(-ikx_0))\exp(ikx)\nonumber\\
&\equiv&u_k'\exp(ikx).
\eea
Now we calculate the Berry's phase of the aformentioned procedure:
\bea 
\theta'_{\rm Berry}&=&\int^{\pi/a}_{-\pi/a}i\bra{u_k'}\frac{\d}{\d k}\ket{u_k'}{\rm d}k\nonumber\\
&=&\theta_{\rm Berry}+2\pi\left(\frac{x_0}{a}\right).
\eea

The shifted wave function thus has a different Berry's phase. This seems to suggest that our entire formalism is wrong, as a rigid translation does not alter the nature of the wave function. 

To understand where the problem comes from, we recall that in order to define the Berry's phase, we need to define a definite way to identify the gauge-equivalent states. Consider the same wave function $\psi_k$, under the large gauge transform of winding number one, it becomes
\bea
\psi_k(x)\rightarrow\bar\psi_k(x)&=&\psi(x)\exp(-i2\pi x/L)\nonumber\\
&=&u_k\exp(i(k-2\pi/L)x);
\eea 
$L$ is the size of the lattice.
Now consider the gauge transform of the shifted wave function $\psi'(x)$:
\bea
\psi_k'(x)\rightarrow\bar\psi_k'(x)&=&u_k'\exp(i(k-2\pi/L)x)\\
&=&u_k(x-x_0)e^{-i2\pi x_0/L}e^{i(k-2\pi/L)(x-x_0)}.\nonumber
\eea
If we identify the wave functions and their large gauge transform (that is, we require that $\ket\psi$ and $\ket{\bar\psi}$ describes the same physical state), then the translated wave functions are identified with the translated large gauge transform of the original wave function with an extra phase $(2\pi x_0/L)$ as shown below in $[\; ]$: (the '$\sim$' symbol here means describing the same physical state.)
\bea
\braket{x}{\psi}&\sim&\braket{x}{\bar\psi}\Rightarrow u_ke^{ikx}\sim u_ke^{i(k-2\pi/L)x}\nonumber\\
\braket{x}{\psi'}&\sim&\braket{x}{\bar\psi'}\\
&\Rightarrow&u_k'e^{ik(x-x_0)}\sim u_k'e^{i(k-2\pi/L)(x-x_0)}[e^{-i2\pi x_0/L}]\nonumber.
\eea

This arises from the fact that the gauge transform does not commute with translation. Once we sum over all occupied states (there are $N\equiv L/a$ of them), the total phase difference between identifying translated wave functions becomes $(2\pi Nx_0/L)=(2\pi x_0/a)$, which is exactly the extra Berry's phase we have picked up. Even though we do the calculation on the set of Bloch wave functions which represent the insulating band electrons, the same result can be obtained for the ions, independent of whether they are bosons or fermions.

The discussion above shows that the Berry's phase for a single charged wave function is not a physical quantity. It depends on how one identifies the wave functions related by a large gauge transform; but for a given identification, the wave functions are identified differently when they are translated. Fortunately for a charge-neutral system, following the same argument, the total Berry's phase is invariant under the translation of the whole system. This total Berry's phase is thus physical and will determine the ground state electric field. However, if we translate either only the electronic wave functions or the ions, the Berry's phase will change accordingly. The Berry's phase, or the coefficient of the $\theta$-term, is thus not determined only by the "topology" of the occupied bands, but also reflects their relative position to the ionic lattice.

Let us now return back to the original problem. In the spinless case, the ions should have the same density as the electrons, which is half a charge per unit cell.  If the ions are localized, they would have a $2a$ period. For the two degenerate ground state, the ionic states are related by a shifted of $a$. Now that we know that a half-period shift of the ions will also give a Berry's phase differed by $\pi$, the total Berry's phase is indeed the same for the two ground states. 

One might wonder how this argument apply for a jellium-like ionic state. The translated ions can look very similar to the original state, and it seems paradoxical for them to have such different Berry's phases. Here we argue that, despite the similarity in the density profile, since we only have one ion per two lattice spacing, the translated state is always very different from the original state, as long as the ions are localized. This is most evident when we look from the single-particle perspective. The center-of-mass positions of the ions must differ by $2a$, and the product wave function is different if we shift it by $a$. One can also imagine the opposite (unphysical) limit, where the ions are delocalized and are described by plane waves. The ionic wave function is then $a$-periodic, but the state becomes gapless, and the Berry's phase procedure does not apply. We thus conclude that for an inert ionic lattice with one ion per two lattice spacing, it can only be $2a$-periodic, and a translation of $a$ gives a different state, with a Berry's phase differed by $\pi$.

\begin{figure}[htb]
	\centering
	\subfigure[]{\includegraphics[width=4cm]{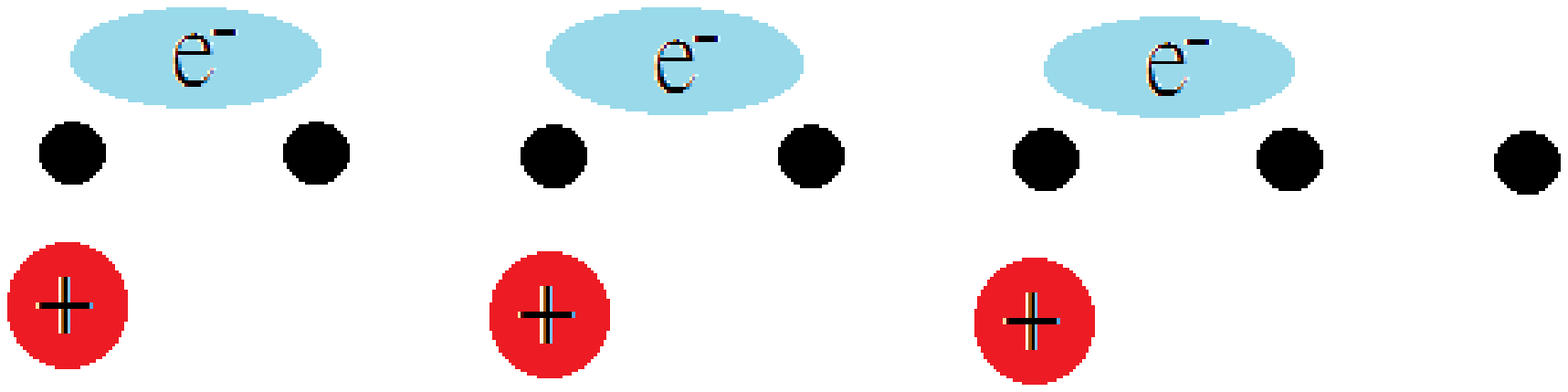}}
	\subfigure[]{\includegraphics[width=4cm]{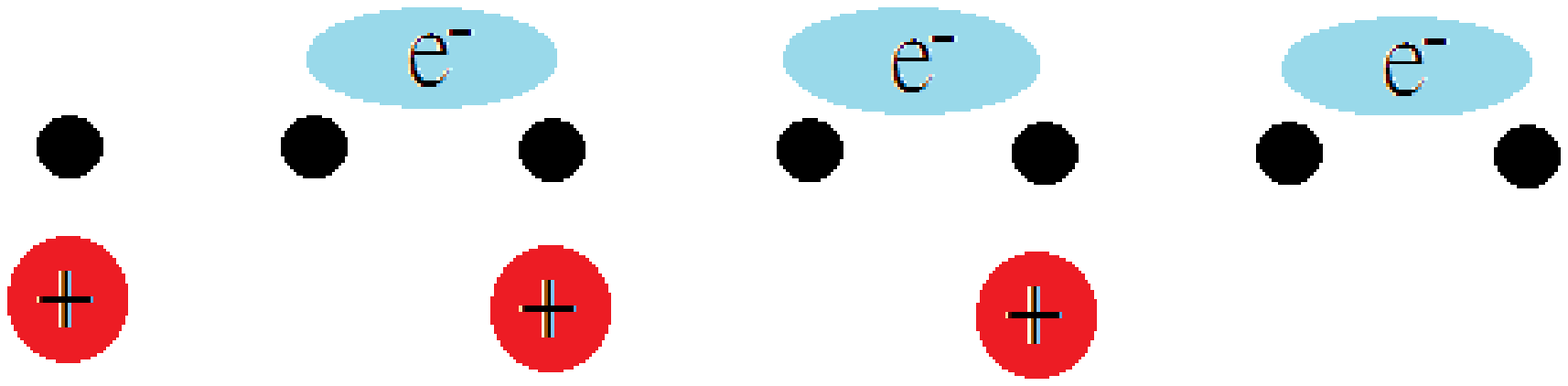}}
	\subfigure[]{\includegraphics[width=4cm]{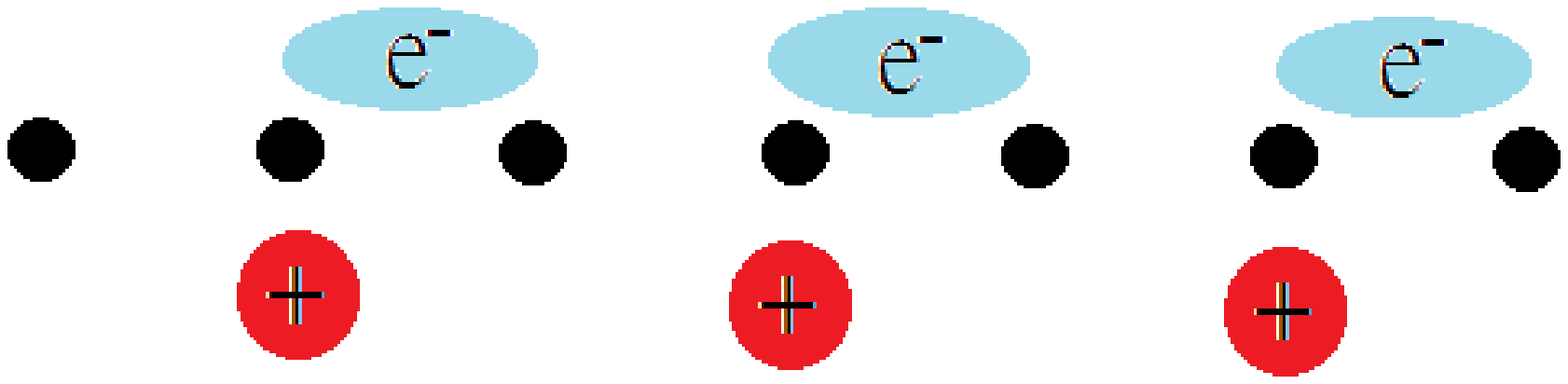}}
	\subfigure[]{\includegraphics[width=4cm]{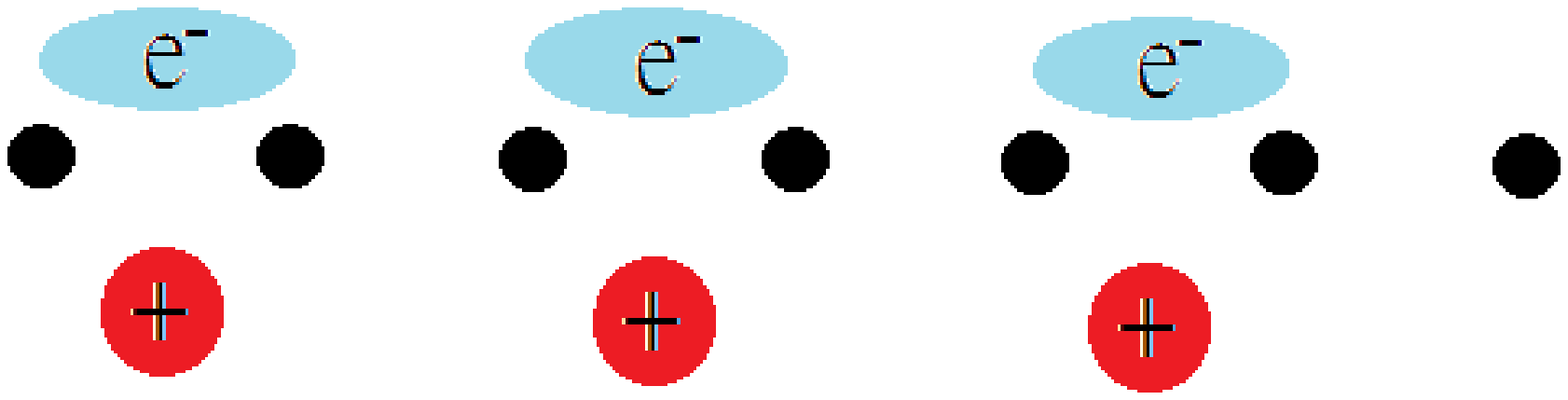}}
	\subfigure[]{\includegraphics[width=4cm]{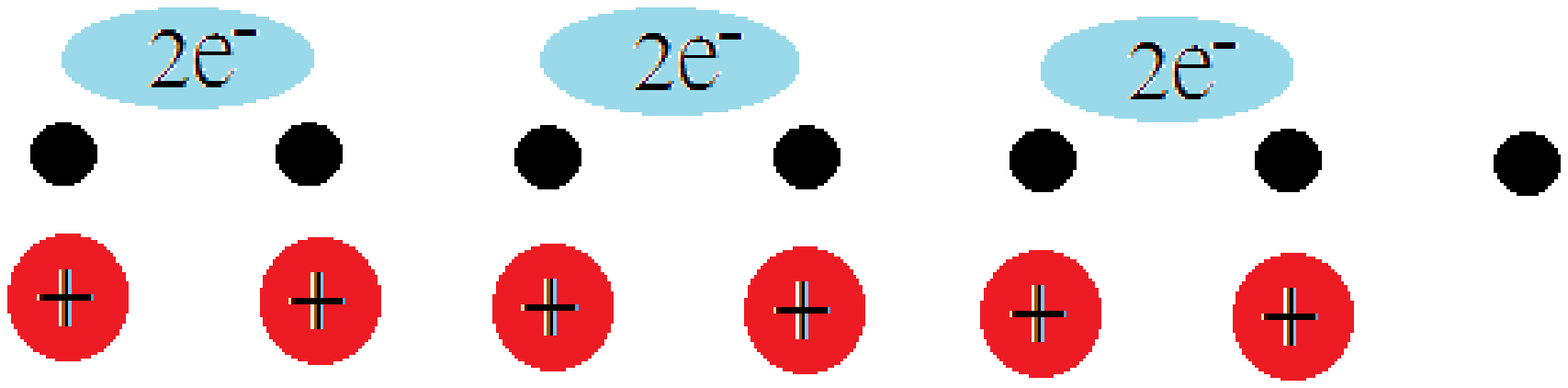}}

	\caption{(a) One of the electronic "ground state", without considering the ions. It is predicted in this state there is a -e/4 static electric field. (b) The other "ground state", without changing the lattice. Evidently the two state are different. (c) The other ground state with ions shifted. Now the physics is identical to (a). Even though we draw point-like ions here, the argument actually works for any charge distribution, including jellium as a limiting case. (d) If the ions are at the lowest energy positions, the ground state electric field is zero. (e) For the spinful case, a simple consideration would show that this configuration will have zero ground state electric field.}
	\label{onlyfig}
\end{figure}

When we derive Eq.(\ref{what}), it is as if we implicitly assume the ions are setting right at $x_n=2na$ (so that they do not contribute to the Berry's phase.) If we place the ions at the places where most electrons are, $x_n={\rm sgn}(\Delta)\frac12a+2na$, the total Berry's phase for both ground states are zero. Fig. (\ref{onlyfig}) summarizes the result.

As for the spinful case, since the number of ions are doubled, the difference between the Berry's phases of the two states is also doubled. The lattice contribution for the two states therefore differs by $2\pi$, which implies that shifting the lattice by $a$ does not change the ground state property. The $\pi$ Berry's phase we obtained does not include the ionic contribution, which is equivalent to assuming they are placed at $x_n=2na$, with two ions at the same site. If we shift half of the ions by $a$, forming the usual lattice with period $a$, the total Berry's phase will again be shifted by $\pi$, and there will be no ground state electric field.

\section{The Dielectric Screening}
Let us start from an effective theory with a dielectric constant:
\beq
\mathcal{L}_{1D}=-\frac\epsilon 4 (F_{\mu\nu})^2+\frac{e\theta}{2\pi}\epsilon^{\mu\nu} \partial_\mu A_\nu=\frac\ep 2E^2+\frac{e\theta}{2\pi}E. 
\eeq
Let us again write down the $q=0$ sector of the partition function following our previous paper:
\begin{widetext}
\beq
Z_{q=0}
\propto
\int_0^{2\pi} \rm d\phi\int^\infty_{-\infty} \frac{\rm d\ell}{2\pi}\sum_{m,n}\braket{\phi+2\pi m}{\ell}\bra{\ell}\exp(-\frac{\beta Le^2}{2\ep}\ell^2)\ket\ell\braket{\ell}{\phi+2\pi n}e^{i(m-n)\theta};
\eeq
\end{widetext}
again, $\phi$ is the initial value of $(e\tilde{A^1}(q=0))$. Note that we now choose $\ell$ to be the eigenvalue of the operator $(\epsilon\tilde{E^1}(q=0)/eL)$, hence the factor of $\ep$ in the denominator of the exponent. Notice that with the modified Lagrangian, it is now $(\epsilon\tilde{E^1}(q=0)/eL)$ which is conjugate to $(e\tilde{A^1}(q=0))$. Therefore, 
\beq
\braket{\phi+2\pi m}{\ell}=\exp(i(\phi+2\pi m)\ell)
\eeq
remain unchanged.

Now we can follow through the same calculation, realizing that \textit{it is $\ell$ that is quantized.} The ground state electric field, following the same argument, should instead be
\beq\label{escreen}
E=-\frac{\theta e}{2\pi\ep},\,\,-\pi<\theta<\pi.
\eeq
This indeed matches the situation with open ends. The ground state electric field is thus not universal.

Up to now this is just an identical calculation as we did in the original paper (but there we overlooked the change of conjugation relation.) Interestingly, one can also understand the screening effect by a shift in $\theta$. In the last section, we have found that $\theta$ shifts by $2\pi$ as we shift the electronic wavefunction by a lattice period. It is thus intuitive to think, that the electrons will shift a little bit, responding to the electric field generated from the $\theta$-term, and make $\theta$ smaller. Here we are going to show that this intuitive picture gives precisely the same effect as above.

From the point of view of the charges, $\theta$ comes from the Berry's phase of the procedure mentioned in the previous section. In the adiabatic limit, we derive that the phase is just the topological index. However, since the $\theta$-term in turn predicts that there in a finite electric field in the ground state, there can be some extra phases coming from the nonadiabaticness.

From the effective theory point of view, the accumulated phase in the presence of a finite field is just the first derivative of the electronic action with respect to the electric field. This gives
\beq
\theta_{{\rm Berry}}=\theta+\frac{2\pi}{e}(\ep-1) E.
\eeq
One can also directly demonstrate this by calculating the accumulated phase of the aformentioned procedure to second order in $E$. We then proceed with the original quantization of the gauge field with this modified $\theta_{\rm Berry}$, we get 
\beq
E=-\frac{\theta_{\rm Berry}e}{2\pi}=-\frac{\theta e}{2\pi}-(\ep-1)E,
\eeq  
and we recover the same result as Eq. (\ref{escreen}). This calculation matches our intuition that the wave function can adjust itself a little bit (a compromise between a rigid shift and the ionic potential, characterized by the dielectric constant $\ep$) to reduce the electric field.

Despite that the two calculations produce the same prediction for the ground state electric field, the quantization of the electric field is evidently different. Ultimately the former calculation is more correct. This stems from the fact that the dynamical accumulated phase is not distinguishable from the energy of the electric system in the presence of the electric field. When one considers the fluctuation of the gauge field to derive the quantization, the dynamical phase will not be proportional to the winding number, and therefore should be thought of as an energy correction. specifically, in the thermal partition function, it should give a real weight instead of a phase. Nevertheless, one can still expect that treating it as a phase should give the same ground state properties. Physically this is because in the ground state the partition function is dominated by the average electric field, and the dynamical phase is proportional to the winding number. There is no way to separate the contribution from the geometric Berry's phase and the dynamical phase in the ground state, since the static electric field is not a free parameter and is determined self-consistently as described above. Technically, the ground state property can be derived from the real time path integral, where there is no real difference between an energy and a phase.

\section{Summary}
In this report, we clarified two issues about the ground state electric field in 1D topological insulators.

 We showed that in a unit-cell doubled system, it is possible for two states related by a lattice translation, to have a different "topological index" characterizing the electronic band structure. It is still a topological index in the sense that we cannot smoothly change from one state to another without breaking the discrete symmetry (in the SSH model, the parity symmetry.) However, since the translation of ions also changes the Berry's phase, the two states are physically equivalent. With the ions placed properly, there will be no $\theta$-term in the effective theory, and no ground state electric field. For a tight-binding model without unit-cell doubling and when the electron orbitals are always tied to to the ions, such as the topological insulator in 1D defined under charge conjugation, the topological index of the electronic band does give a ground state electric field, and the topological state and the trivial state are intrinsically different.

We also showed that, unlike previously stated, the electric field is not perfectly quantized in a system without boundaries. The screening effect, can be either viewed as a change of the quantization of the static electric field in the presence of the dielectric constant, or as a shift of $\theta$ in the presence of the field it generates.

We thank C. Kane for pointing out the issues in the SSH model and the fruitful discussion. We acknowledge the support of NSF under grant DMR 0804040.


\end{document}